\documentclass[preprint]{ptephy_v1}%%%%%% to generate preprint number with ptep logo

\preprintnumber{XXXX-XXXX} %%% %%% Insert preprint number here
\usepackage{hyperref}
\usepackage{ulem}
\usepackage{color}
\begin{document}

\title{Proper derivation of subspace mapping from whole space mapping in boson expansion theory}

%%%% To generate auto affiliation numbers please use \author{}\affil{} command

\author{Kimikazu Taniguchi}
\affil{Department of Health Data Science, Suzuka University of Medical Science, Suzuka  1001-1, Japan\email{kmkzt01@gmail.com}}

%\author{Insert second author name here}
%\affil{Insert second author address here}

%\author{Insert third author name here}
%\author[3]{Insert fourth author name here} %%% Use optional bracket [3] to change the respective address
%\affil{Insert third author address here}

%\author{Insert last author name here\thanks{These authors contributed equally to this work}}
%\affil{Insert last author address here}

%%% To include the collaborator name... Please use the command "\collaborator"
%%% For example: \collaborator{ATLAS Collaboration}

\begin{abstract}
The norm operator method, which was recently proposed as a new formulation of the boson expansion theory (BET), is used to show that the subspace mapping is properly derived from the whole space mapping. This derivation requires the appropriate renormalization of the contribution of phonons that are not adopted as boson excitations in the subspace mapping. This was impossible with conventional BETs (which ignore these contributions), and is only made possible for the first time by the norm operator method, which treats these contributions appropriately. We also correct the confusion in the claims of conventional BETs.
Namely, contrary to conventional claims, we show that when the phonon excitations not adopted as boson excitations make no contribution at all, the subspace mapping is obtained simply by discarding those excitations. Furthermore, we demonstrate that the Park operator, which had been considered effective only in the whole space mapping, is also effective in the subspace mapping.
These findings provide a clear criterion for verifying the applicability of the boson expansion theory to large-amplitude collective motions and offer a new perspective on a microscopic foundation of the interacting boson model (IBM).
\end{abstract}

\subjectindex{xxxx, xxx}

\maketitle

\section{Introduction}
Elucidating the microscopic structure of finite quantum many-body systems such as atomic nuclei remains a challenging task.
The boson expansion theory (BET) is one of the many-body theories used to elucidate the large amplitude collective motions of atomic nuclei \cite{KM91}.

The boson expansion theory can be formulated by the one-to-one mappings of the fermion space (which consists of an even number of quasi-particles or its subspace) into the boson subspace \cite{Usi60, MYT64,  JDF71, LH75, KT83, Ta01}.
These mappings embed the effect of the Pauli exclusion principle into the boson subspace. The boson operators obtained by mapping are expressed in an expansion form, which embodies the effect of the Pauli exclusion principle, and the boson state vectors that correspond to the fermion ones also generally reflect this effect.

The whole fermion space mapping (WFSM) introduces a boson operator corresponding to each quasi-particle pair operator. The fermion space that is composed of these quasi-particle pairs is then mapped onto a boson subspace constructed from these boson operators. The mapping operator is obtained by utilizing a one-to-one correspondence between basis states containing an even number of quasi-particles and the completely antisymmetrized boson state vectors \cite{MYT64}.
This method faithfully reproduces the original whole fermion space within the boson subspace. However, the convergence of the boson expansion in this method is poor (with the exception of the Dyson boson expansion \cite{JDF71}, which results in a finite expansion), and the state vectors to be used become complicated, rendering it impractical for applications.

To obtain a practical boson expansion method, a method was developed that only takes the collective excitation modes of the Tamm-Dancoff type phonons into account \cite{LH75}.
This method is a modified version of the Marumori, Yamamura, and Tokunaga (MYT) method \cite{MYT64}. It is formulated as a method of mapping a fermion space (which is pre-restricted to a subspace spanned by multi-phonon state vectors composed of collective excitation modes) onto a boson subspace composed only of excitation modes corresponding to those phonon modes.
To obtain the mapping operator, the multi-phonon state vectors (composed of collective excitation modes) must be normalized in advance and matched with the corresponding normalized multi-boson state vectors. It is emphasized that the mapping operator cannot be obtained simply by rewriting the WFSM's mapping operator with Tamm-Dancoff type phonons and discarding the non-adopted modes.
This means that the fermion space must be restricted in advance, rather than mapping the whole fermion space onto the boson subspace and then restricting the excitation modes later.
This method has been extended to incorporate some non-collective excitation modes and has been adopted as the basis for the KT-3 \cite{KT83} and the Dyson boson expansion theory \cite{Ta01}.
The mapping operator is constructed by a one-to-one correspondence between the orthonormalized multi-phonon state vectors (composed of collective and some non-collective excitation modes) and the multi-boson state vectors composed only of phonons corresponding to the adopted phonon operators.
We refer to this as the Fermion Subspace Mapping (FSSM).

Since the mapping operator cannot be obtained, the mapping results of the phonon and scattering operators via FSSM should also not be obtainable simply by adding restrictions on the boson excitation modes to those obtained via WFSM.
In fact, it is emphasized that the boson expansion of KT-3 cannot be obtained by this method \cite{KT83, SK88}.
However, in the Dyson boson expansion (DBE) theory, attention is drawn to the fact that the boson expansion obtained by FSSM can be reproduced by applying the DBE of quasi-particle pairs (obtained via WFSM) to the phonon operators and scattering operators, and then limiting the boson excitation modes.
Furthermore, while emphasizing the preservation of the algebraic structure between the results obtained from WFSM and FSSM (by adopting the name "skeleton boson realizations" \cite{KV88}), it is argued that the Park operator \cite{P87} (which is used to examine whether a boson state vector in WFSM is physical, i.e., has a counterpart in the fermion space) cannot be used in FSSM, despite being derived by the DBE theory.
They argue that the difficulty concerning the Park operator can be avoided by attributing the cause to the difference in the projection operators onto the physical subspace in WFSM and FSSM.

The cause of this confusion may originate from the formulation of the conventional FSSM, where the contribution of phonons with excitation modes not adopted as boson excitations is discarded (Non-Adopted-Mode-Discretion: NAMD) \cite{HJJ76, KT83, Ta01}.
The difference between WFSM and FSSM lies in the fact that the former treats all quasi-particle pair modes as boson excitations, while the latter treats only some excitation modes (including the collective excitation modes of phonons) as boson excitations.
The physical subspace obtained via WFSM is a complete replica of the whole fermion space. Therefore, the physical subspace of FSSM should be obtained by appropriately renormalizing the excitation modes not adopted as boson excitations in FSSM into that physical subspace.
However, in the conventional boson expansion theory, the excitations other than those adopted in the latter (FSSM) are not considered at all when deriving the FSSM physical subspace from the WFSM, corresponding to NAMD; FSSM is discussed solely in the context of excluding all non-adopted modes.
Furthermore, while the conventional boson expansion theory adopts NAMD as a "good approximation," it cannot provide a clear justification for this, as NAMD is a prerequisite for its formulation.

In recent years, a new boson expansion theory that does not presuppose NAMD, the norm operator method \cite{NR23}, has been proposed.
This method introduces a mapping that limits not only the types of phonon excitation modes (which were conventionally adopted) but also the number of excitations to obtain a practical boson expansion.
By introducing the norm operator into this mapping operator, it is possible to handle Hermitian and non-Hermitian type mappings comprehensively. Furthermore, by removing the limitation imposed on the phonon excitations, both the whole fermion space mapping and the conventionally adopted fermion subspace mapping can be reproduced.
The introduction and utilization of the norm operator allowed us to establish a concrete method for the small parameter expansion (where the phonon becomes a boson in the zeroth approximation), and enabled an accurate understanding of cases where NAMD holds.
Namely, the following points have been clarified:
\begin{enumerate}
\item
The small parameter expansion and NAMD are incompatible.
\item
If the small parameter expansion holds, the expansion becomes an infinite expansion regardless of whether it is a Hermitian or non-Hermitian type.
\item
If NAMD holds, the expansion is effectively a finite expansion.
\end{enumerate}
These findings overturn the conclusions of conventional boson expansion theories. The results obtained by conventional boson expansion theories need to be re-examined using this norm operator method.

In this paper, we review the claims of conventional boson expansion theories regarding the relationship between WFSM and FSSM, and re-examine them using the norm operator method.

Section \ref{cbet} deals with conventional boson expansion theories, focusing on both WFSM and FSSM. In particular, we emphasize that the mapping becomes an isomorphic mapping in WFSM, whereas an isomorphic mapping can be obtained in FSSM only under appropriate assumptions.
An isomorphic mapping guarantees that the mapping of quasi-particle pair operators preserves the original commutation relations.
Furthermore, we point out that the relationship between WFSM and FSSM, as discussed by conventional boson expansion theories, has been inadequately addressed, leading to confusion.

Section \ref{normac} provides an analysis using the norm operator and presents the conclusions. First, after describing the outline of the norm operator method, we mention the results obtained so far. Then, we analyze the relationship between WFSM and FSSM using the norm operator method, establish a method for deriving the physical subspace of FSSM from the physical subspace of WFSM, and correct the errors in the claims of conventional boson expansions.

Section \ref{dis} discusses how to verify the applicability of the boson expansion theory to large-amplitude collective motions and offers a new perspective on a microscopic foundation of the interacting boson model (IBM).

Section \ref{sum} provides the summary of this paper.

\section{Conventional Boson Expansion Theories}\label{cbet} \subsection{Whole Fermion Space Mapping} The whole fermion space mapping (WFSM) introduces boson creation and annihilation operators \(b_{\alpha \beta }^{\dag }\) and \(b_{\alpha \beta }\) (\(b_{\beta \alpha }=-b_{\alpha \beta }\)) that correspond to all quasi-particle pair creation and annihilation operators \(a_{\alpha }^{\dag }a_{\beta }^{\dag }\) and \(a_{\beta }a_{\alpha }\), respectively. This maps the entire fermion space consisting of an even number of quasi-particles one-to-one onto a boson subspace.

The quasi-particle creation and annihilation operators satisfy the following commutation relations:
\begin{equation}
\begin{array}{lll}
[a_{\beta'}a_{\alpha'}, a_\alpha^\dagger a_\beta^\dagger]
&=&\delta_{\alpha'\alpha}\delta_{\beta'\beta}-\delta_{\alpha'\beta}\delta_{\beta'\alpha}
\\
&&
-\delta_{\alpha'\alpha}a_\beta^\dagger a_{\beta'}+\delta_{\alpha'\beta}a_\alpha a_{\beta'}+\delta_{\beta'\alpha}a_\beta^\dagger a_{\alpha'}
\\
&&-\delta_{\beta'\beta}a_\alpha^\dagger a_{\alpha'}+\delta_{\beta'\beta}a_\alpha^\dagger a_{\alpha'}
+a_\alpha^\dagger a_\beta^\dagger a_{\beta'}a_{\alpha'}\text{,}
\end{array}
\end{equation}
whereas the commutation relations for the boson creation and annihilation operators are as follows:
\begin{equation}
[b_{\alpha'\beta'}, b_{\alpha\beta}^\dagger]=\delta_{\alpha'\alpha}\delta_{\beta'\beta}-\delta_{\alpha'\beta}\delta_{\beta'\alpha}\text{.}
\end{equation}

The mapping operator from the whole fermion space to the boson space \cite{MYT64} is constructed by establishing a one-to-one correspondence between \(|\alpha _{1}\beta _{1}\cdots \alpha _{N}\beta _{N}\rangle =a_{\alpha _{1}}^{\dag }a_{\beta _{1}}^{\dag }\cdots a_{\alpha _{N}}^{\dag }a_{\beta _{N}}^{\dag }|0\rangle \) and \(|\alpha _{1}\beta _{1}\cdots \alpha _{N}\beta _{N})_{\mathcal{A}}\), which is obtained by normalizing and completely antisymmetrizing the state \(b_{\alpha _{1}\beta _{1}}^{\dag }\cdots b_{\alpha _{N}\beta _{N}}|0)\) with respect to the indices. 
The mapping operator
\begin{equation}
U_H^{(W)}=\sum_{N=0}^\infty\sum_{\alpha_1\beta_1<\cdots<\alpha_N\beta_N}\vert \alpha_1\beta_1\cdots\alpha_N\beta_N)_{\mathcal A}\langle\alpha_1\beta_1\cdots\alpha_N\beta_N\vert\text{,}
\end{equation}
transforms the quasi-particle pair operator \(O_{F}\) and the state vector \(|\psi \rangle \) in the whole fermion space as follows:
\begin{subequations}
\begin{equation}
\begin{array}{lll}
(O_F)_H^{(W)}&=&U_H^{(W)}O_F\left\{U_H^{(W)}\right\}^\dagger;\quad (O_F^\dagger)_H^{(W)}=\left\{(O_F)_H^{(W)}\right\}^\dagger\text{,}
\\
(O_FO'_F)_H^{(W)}&=&(O_F)_H^{(W)}(O'_F)_H^{(W)}\text{,}
\end{array}
\end{equation}
\begin{equation}
\vert\psi)_H^{(W)}=U_H^{(W)}\vert\psi\rangle;\quad {}^{(W)}_{\quad H}(\psi\vert=\left\{\vert\psi)_H^{(W)}\right\}^\dagger\text{.}
\end{equation}
\end{subequations}
This is an isomorphic mapping from the whole space to the physical subspace that preserves the original Hermitian conjugation. 
The mapping operator also satisfies the following relations:
\begin{subequations}\begin{equation}U_H U_H^\dagger =\hat T_B^{(W)};\quad \hat T_B^{(W)}=\sum_{N=0}^\infty\sum_{\alpha_1\beta_1<\cdots<\alpha_N\beta_N}\vert \alpha_1\beta_1\cdots\alpha_N\beta_N)_{\mathcal A}{}_{\mathcal A}\langle\alpha_1\beta_1\cdots\alpha_N\beta_N\vert\text{,}
\end{equation}
\begin{equation}
U_H^\dagger U_H =\hat 1_B;\quad \hat 1_B=\sum_{N=0}^\infty\sum_{\alpha_1\beta_1<\cdots<\alpha_N\beta_N}\vert \alpha_1\beta_1\cdots\alpha_N\beta_N\rangle\langle\alpha_1\beta_1\cdots\alpha_N\beta_N\vert\text{.}
\end{equation}\end{subequations}
Since $\hat {T}_{B}^{(W)}$ is the projection operator onto the physical subspace, and $\hat {1}_{B}$ is the identity operator of the entire space consisting of quasi-particle pairs, this mapping guarantees that
\begin{equation}
{}^{(W)}_{\quad H}(\psi'\vert (O_F)_H\vert\psi)_H^{(W)}=\langle\psi'\vert O_F\vert\psi\rangle\text{.}
\end{equation}

The mapping operator is not limited to this form \cite{JDF71}. For instance, operators defined as
\begin{equation}
\begin{array}{c}
U_1^{(W)}=\sqrt{(2\hat{N}^{(W)}_B-1)!!}U_H,\quad U_2^{(W)}=\displaystyle\frac{1}{\sqrt{(2\hat{N}^{(W)}_B-1)!!}}U_H
\\
\hat{N}_B^{(W)}=\displaystyle\sum_{\alpha<\beta}b_{\alpha\beta}^\dagger b_{\alpha\beta}
\end{array}
\end{equation}
satisfy the relations
\begin{equation}
U_1^{(W)}\left\{ U_2^{(W)}\right\}^\dagger =\hat{T}_B^{(W)},\quad
\left\{U_2^{(W)}\right\}^\dagger U_1^{(W)} =\hat{1}_B\text{,}
\end{equation}
and the transformation is given by
\begin{subequations}
\begin{equation}
(O_F)_{NH}^{(W)}=U_1^{(W)}O_F\left\{U_2^{(W)}\right\}^\dagger;\quad (O_F O'_F)_{NH}^{(W)}=(O_F)_{NH}^{(W)}(O'_F)_{NH}^{(W)}\text{,}
\end{equation}
\begin{equation}
\vert\psi)_R^{(W)}=U_1^{(W)}\vert\psi\rangle,\quad {}^{(W)}{}_L(\psi\vert=\langle\psi\vert\left\{U_2^{(W)}\right\}^\dagger\text{.}
\end{equation}
\end{subequations}
This transformation is also an isomorphic mapping from the whole space to the physical subspace, but it is clear that it does not preserve the original Hermitian conjugation:
\begin{equation}
(O_F^\dagger)_{NH}^{(W)}\neq\left\{ (O_F)_{NH}^{(W)}\right\}^\dagger,\quad {}^{(W)}{}_L(\psi\vert\neq\vert\psi)_R^{(W)}\text{.}
\end{equation}

The mapping of quasi-particle pairs is given as follows \cite{JDF71}:
\begin{subequations}
\begin{equation}
\begin{array}{lll}
(a_\beta a_\alpha)_H&=&\left\{(a_\alpha^\dagger a_\beta^\dagger)_D^{(W)}\right\}^\dagger\displaystyle\frac{1}{\sqrt{1+2\hat{N}_B^{(W)}}}\hat{T}_B^{(W)}\text{,}
\\
&=&(a_\beta a_\alpha)_{HP}\hat{T}_B^{(W)}\text{,}
\\
(a_\beta^\dagger a_\alpha)_H&=&(a_\beta^\dagger a_\alpha)_{HP}\hat{T}_B^{(W)}\text{.}
\end{array}
\end{equation}
\begin{equation}
\begin{array}{lll}
(a_\beta a_\alpha)_{NH}^{(W)}&=&(a\beta a_\alpha)_D^{(W)}\hat{T}_B^{(W)}\text{,}
\\
(a_\alpha^\dagger a_\beta^\dagger)_{NH}^{(W)}&=&(a_\alpha^\dagger a_\beta^\dagger)_D^{(W)}\hat{T}_B^{(W)}\text{;}
\\
(a_\beta^\dagger a_\alpha)_{NH}^{(W)}&=&(a_\beta^\dagger a_\alpha)_D^{(W)}\hat{T}_B^{(W)}\text{.}
\end{array}
\end{equation}
\end{subequations}
Here,
\begin{equation}
\begin{array}{lll}
(a_\beta a_\alpha)_{HP}&=&\left (\sqrt{1-\rho}\ b\right )_{\alpha\beta};\quad\hat \rho_{\alpha\beta}=\sum_\gamma b_{\beta\gamma}^\dagger b_{\alpha\gamma}\text{,}
\\
(a_\beta^\dagger a_\alpha)_{HP}&=&\sum_\gamma b_{\beta\gamma}^\dagger b_{\alpha\gamma}\text{,}
\end{array}
\end{equation}
are the Holstein-Primakoff boson expansions summed up to infinite terms, and
\begin{equation}
\begin{array}{lll}
(a_\beta a_\alpha)_D^{(W)}&=&b_{\alpha\beta}\text{,}
\\
 (a_\alpha^\dagger a_\beta^\dagger)_D^{(W)}&=&b_{\alpha\beta}^\dagger-\sum_{\gamma\delta}b_{\alpha\gamma}^\dagger b_{\beta\gamma}^\dagger b_{\gamma\delta}\text{,}
\\
(a_\beta^\dagger a_\alpha)_D^{(W)}&=&\sum_\gamma b_{\beta\gamma}^\dagger b_{\alpha\gamma}\text{,}
\end{array}
\end{equation}
are the Dyson boson expansions.
From these results, if we assume $O_F$ to be a quasi-particle pair operator, then for the physical state vectors, which are eigenvectors of $\hat T_B$ with an eigenvalue of one, the following relations hold:
\begin{subequations}
\begin{equation}
{}_H(\psi'\vert (O_F)_H\vert\psi)_H={}_H(\psi'\vert (O_F)_{HP}\vert\psi)_H\text{,}
\end{equation}
\begin{equation}
{}^{(W)}_{\quad L}(\psi'\vert (O_F)_{HN}^{(W)}\vert\psi)_R^{(W)}={}^{(W)}_{\quad L}(\psi'\vert (O_F)_D^{(W)}\vert\psi)_R^{(W)}\text{.}
\end{equation}
\end{subequations}
This implies that the mapping of quasi-particle pair operators can be accomplished simply by obtaining the Holstein-Primakoff and Dyson boson expansions, respectively.
On the other hand, Marumori, Yamamura, and Tokunaga (MYT) \cite{MYT64} derived the boson expansion of the quasi-particle pair operator by expressing the projection operator onto the boson vacuum, $\vert 0)(0\vert$, in an expansion form. However, this method has a defect in that the expansion of $\vert 0)(0\vert$ is not a small parameter expansion and has poor convergence; higher-order terms are required as the number of boson excitations treated increases, as shown in the following expansion:
\begin{equation}
(a_\beta a_\alpha)_H^{(W)}=b_{\alpha\beta}-\left (1-\frac{1}{\sqrt 3}\right )\hat N_B^{(W)}b_{\alpha\beta}-\frac{1}{\sqrt 3}\sum_{\gamma\delta}b_{\gamma\delta}b_{\beta\delta}b_{\alpha\gamma}+\cdots\text{.}
\end{equation}

While the completely antisymmetrized state vectors $\vert \alpha_1\beta_1\cdots\alpha_N\beta_N)_{\mathcal A}$ are physical state vectors, a general physical state vector obtained by their superposition is generally complicated, and it is not always self-evident whether it can be expressed in a simple form. Historically, the boson expansion of quasiparticle pair operators was constructed to reproduce the commutation relations of the quasiparticle pairs, but it lacked consideration for the physical state vectors. With this background, in the whole space mapping, an operator was devised to determine whether a state vector in the boson space is physical. One such operator is the Park operator \cite{P87}.

The definition of the Park operator is as follows:
\begin{equation}
\hat S=\{(\hat N_F^2)_D^{(W)}\}^{dir}-\{(\hat N_F^2)_D^{(W)}\}^{ex}\text{,}
\end{equation}
\begin{subequations}
where
\begin{equation}
\{(\hat N_F^2)_D^{(W)}\}^{dir}=\{(\hat N_F)_D^{(W)}\}^2; \quad
\hat N_F=\sum_\alpha a_\alpha^\dagger a_\alpha\text{,}
\end{equation}
\begin{equation}
\{(\hat N_F^2)_D^{(W)}\}^{ex}=(\hat N_F)_D^{(W)}
-\sum_{\alpha\beta}(a_\alpha^\dagger a_\beta^\dagger)_D^{(W)}(a_\alpha a_\beta)_D^{(W)}\text{,}
\end{equation}
\begin{equation}
\begin{array}{lll}
(a_\alpha^\dagger a_\beta^\dagger)_D^{(W)}&=&b_{\alpha\beta}^\dagger -\displaystyle\sum_{\gamma\delta}b_{\alpha\gamma}^\dagger b_{\beta\gamma}^\dagger b_{\gamma\delta}\text{,}
\\
(a_\beta a_\alpha)_D^{(W)}&=&b_{\alpha\beta}\text{,}
\\
(a_\alpha^\dagger a_\beta)_D^{(W)}&=&\displaystyle\sum_\gamma b_{\alpha\gamma}^\dagger b_{\beta\gamma}\text{.}
\end{array}
\end{equation}
\end{subequations}
If we assume that $\vert phys)$ is a physical state vector, then $\hat T_B^{(W)}\vert phys)=\vert phys)$ holds. Considering that $(O_F)_{NH}^{(W)}\vert phys)$ is also a physical state vector, and that the mapping is an isomorphic mapping, the following relation holds:
\begin{equation}
\label{eq:oo'}
\begin{array}{lll}
(O_F)_D^{(W)}(O'_F)_D^{(W)}\vert phys)&=&(O_F)_D^{(W)}(O'_F)_{NH}^{(W)}\vert phys)
\\
&=&(O_F)_{HN}^{(W)}(O'_F)_D^{(W)}\vert phys)
\\
&=&(O_FO'_F)_D^{(W)}\vert phys)\text{.}
\end{array}
\end{equation}
From these results, applying the Park operator to the physical state vector yields the following result:
\begin{equation}
\hat S\vert phys)=(\hat N_F^2-\hat N_F-\sum_{\alpha\beta}a_\alpha^\dagger a_\beta^\dagger a_\alpha a_\beta)_D^{(W)}\vert phys)=0\vert phys)\text{.}
\end{equation}
That is, the physical state vectors are the eigenstates of the Park operator with an eigenvalue of zero. On the other hand, for unphysical state vectors, Eq. (\ref{eq:oo'}) does not hold, and they are not eigenvectors of the Park operator with the eigenvalue zero.

\subsection{Fermion Subspace Mapping}

In the case of infinite expansion of quasiparticle pairs, WFSM not only has poor convergence of the expansion but also the physical state vectors become complicated due to the effect of the Pauli exclusion principle.
To improve this, a method has been developed to map only a fermion subspace \cite{LH75, KT83, Ta01}.
It is expected that by limiting the phonons obtained as a superposition of quasiparticle pairs to those carrying collective excitation modes, the effect of the Pauli exclusion principle is diminished, which not only improves the convergence of the boson expansion but also allows for simpler physical state vectors.

Tamm-Dancoff type phonons are adopted as the phonons. Their creation and annihilation operators are as follows:
\begin{equation}
\label{eq:phononopc}
X_\mu^\dagger
=\displaystyle\sum_{\alpha<\beta}\psi_\mu(\alpha\beta)a_\alpha^\dagger
a_\beta^\dagger\text{,}
\quad
X_\mu=\displaystyle\sum_{\alpha<\beta}\psi_\mu(\alpha\beta)a_\beta
a_\alpha\text{.}
\end{equation}
In addition to these, scattering operators are also introduced:
\begin{equation}
\label{eq:scop}
B_q=\sum_{\alpha\beta}\varphi_q(\alpha\beta)a_\beta^\dagger
a_\alpha\text{.}
\end{equation}
The coefficients satisfy the following relations:
\begin{subequations}
\label{eq:TDrel}
\begin{equation}
\label{eq:anti}
\psi_\mu(\beta\alpha)=-\psi_\mu(\alpha\beta)\text{,}
\end{equation}
\begin{equation}
\label{eq:orthonormal}
\sum_{\alpha<\beta}\psi_\mu(\alpha\beta)\psi_{\mu'}(\alpha\beta)=\delta_{\mu,
\mu'}\text{,}
\end{equation}
\begin{equation}
\label{eq:complete}
\sum_{\mu}\psi_\mu(\alpha\beta)\psi_{\mu}(\alpha'\beta')=\delta_{\alpha,
\alpha'}\delta_{\beta, \beta'}-\delta_{\alpha,
\beta'}\delta_{\beta, \alpha'}\text{,}
\end{equation}
\end{subequations}
\begin{subequations}
\label{eq:coeffscop}
\begin{equation}
\label{eq:barq}
\varphi_{\bar q}(\alpha\beta)=\varphi_q(\beta\alpha)\text{.}
\end{equation}
\begin{equation}
\label{eq:coeffscop1}
\displaystyle\sum_{\alpha\beta}\varphi_q(\alpha\beta)\varphi_{q'}(\alpha\beta)=\delta_{q,q'}\text{,}
\end{equation}
\begin{equation}
\label{eq:coeffscop2}
\displaystyle\sum_{q}\varphi_q(\alpha\beta)\varphi_{q}(\alpha'\beta')=\delta_{\alpha,\alpha'}\delta_{\beta,
\beta'}\text{.}
\end{equation}
\end{subequations}
The phonon and scattering operators satisfy the following commutation relations:
\begin{subequations}
\label{eq:algebra}
\begin{equation}
\label{eq:algebra1}
[ X_\mu, X_{\mu'}^\dagger ]=\delta_{\mu,
\mu'}-\sum_q\Gamma^{\mu\mu'}_qB_q\text{,}
\end{equation}
\begin{equation}
\label{eq:algebra2}
[ B_q, X_\mu^\dagger
]=\sum_{\mu'}\Gamma^{\mu\mu'}_qX_{\mu'}^\dagger\text{,}
\end{equation}
\begin{equation}
\label{eq:algebra3}
[ X_\mu, B_q ]=\sum_{\mu'}\Gamma^{\mu'\mu}_qX_{\mu'}\text{.}
\end{equation}
\end{subequations}
where the definition of $\Gamma^{\mu\mu'}_q$ is as follows:
\begin{equation}
\label{eq:Gamma}
\Gamma^{\mu\mu'}_q=\sum_{\alpha\beta}\varphi_q(\alpha\beta)\Gamma^{\mu\mu'}_{\alpha\beta}\text{,}
\quad
\Gamma^{\mu\mu'}_{\alpha\beta}=\sum_\gamma\psi_\mu(\alpha\gamma)\psi_{\mu'}(\beta\gamma)\text{.}
\end{equation}
The following relation holds:
\begin{equation}
\label{eq:qbargam}
\Gamma_{\bar q}^{\mu_1 \mu_2}=\Gamma_q^{\mu_2 \mu_1}\text{.}
\end{equation}
From Eqs. (\ref{eq:algebra1}) and (\ref{eq:algebra2}), we obtain
\begin{equation}
\label{eq:doublecom}
[ [X_{\mu_1}, X_{\mu_2}^\dagger], X_{\mu_3}^\dagger] = -\sum_{\mu'}Y(\mu_1, \mu_2, \mu_3, \mu')X_{\mu'}^\dagger\text{,}
\end{equation}
where the definition of $Y(\mu_1\mu_2\mu_3\mu_4)$ is 
\begin{equation}
\label{eq:Y}
Y(\mu_1\mu_2\mu_3\mu_4)=\sum_q\Gamma_q^{\mu_1\mu_2}\Gamma_q^{\mu_3\mu_4}
=\sum_{\alpha\beta}\Gamma_{\alpha\beta}^{\mu_1\mu_2}\Gamma_{\alpha\beta}^{\mu_3\mu_4}\text{.}
\end{equation}
The following relation holds:
\begin{equation}
\label{eq:Ysym}
\begin{array}{lll}
Y(\mu_1\mu'_1\mu'_2\mu_2)
&=&Y(\mu_2\mu'_1\mu'_2\mu_1)\text{,}
\\
&=&Y(\mu_1\mu'_2\mu'_1\mu_2)\text{,}
\\
&=&Y(\mu'_1\mu_1\mu_2\mu'_2)\text{.}
\\
\end{array}
\end{equation}

To construct the fermion subspace, we divide the entire set of excitation modes $\{\mu\}$ into two sets: $\{t\}$, which includes the collective excitation modes, and $\{\bar t\}$, which includes the others. We then introduce only the boson creation and annihilation operators, $b_t^\dagger$ and $b_t$, corresponding to the phonon operators with the former excitation modes.
The subspace spanned by
\begin{equation}
\vert N; t\rangle =\vert t_1\cdots t_N\rangle={\mathcal N}(N; t)^{-\frac 12}X_{t_1}^\dagger X_{t_2}^\dagger\cdots X_{t_N}^\dagger \vert 0\rangle
\end{equation}
is mapped onto the boson space spanned by the orthonormal system
\begin{equation}
\vert N; t)=\vert t_1\cdots t_N)={\mathcal N}(N; t)^{-\frac 12}b_{t_1}^\dagger\cdots b_{t_N}^\dagger\vert 0)\text{.}
\end{equation}
The state vectors $\vert N;t\rangle$ are generally not an orthonormal system and are overcomplete. To construct the mapping operator, we create an orthonormal system from them. This system is constructed from the orthonormalized eigenvectors of the multi-phonon norm matrix $\langle N; t\vert N; t'\rangle$, excluding those with zero eigenvalues. That is, we utilize
\begin{subequations}
\begin{equation}
\sum_t\langle N; t\vert N; t'\rangle u_a^{t'}(N)=z_a(N)u_a^t(N)\text{,}
\end{equation}
\begin{equation}
\sum_tu_a^t(N)u_{a'}^t=\delta_{a,a'}\text{,}
\end{equation}
\begin{equation}
\sum_a u_a^t(N)u_a^{t'}=\delta_{t,t'}\text{,}
\end{equation}
\end{subequations}
to create the orthonormal system
\begin{equation}
\vert N; a\rangle=z_a(N)^{-\frac 12}\sum_tu_a^t\vert N; t\rangle;\quad (a\neq a_0; a_0: z_{a_0}(N)=0)\text{.}
\end{equation}
Then, to make the orthonormal system in the boson space correspond to $\vert N; a\rangle$, we transform the orthonormal system of the boson space using the same coefficients $u_a^t(N)$ as follows:
\begin{equation}
\vert N; a)=\sum_tu_a^t\vert N;t)\text{.}
\end{equation}
By doing this, a strict one-to-one correspondence is established between $\vert N; a\rangle$ and $\vert N; a)$ for which $a\neq a_0$.

The Hermitian-type mapping operator is given as follows:
\begin{equation}
U_H^{(S)}=\sum_{N=0}^\infty\sum_{a\neq a_0}\vert N; a)\langle N; a\vert\text{.}
\end{equation}
This mapping operator satisfies the following relations:
\begin{subequations}
\begin{equation}
U_H^{(S)}\left\{U_H^{(S)}\right\}^\dagger =\hat T_B^{(S)};\quad \hat T_B^{(S)}=\sum_{N=0}^\infty\sum_{a\neq a_0}\vert N; a)(N; a\vert\text{,}
\end{equation}
\begin{equation}
\left\{U_H^{(S)}\right\}^\dagger U_H^{(S)}=\hat T_F^{(S)};\quad \hat T_F^{(S)}=\sum_{N=0}^\infty\sum_{a\neq a_0}\vert N; a\rangle\langle N; a\vert\text{.}
\end{equation}
\end{subequations}
$\hat T_F^{(S)}$ is the projection operator onto the fermion subspace to be mapped, while $\hat T_B^{(S)}$ is the projection operator onto the physical subspace in the boson space. The mapping operator $U_H^{(S)}$ provides a one-to-one mapping between the subspaces projected by these operators.

In practice, if we let $O_F$ and $\vert\psi\rangle$ be the phonon and scattering operators and state vector of the whole fermion space, respectively, the mapping operator transforms them as follows:
\begin{subequations}
\begin{equation}
(O_F)_H^{(S)}=U_H^{(S)}O_F\left\{U_H^{(S)}\right\}^\dagger;\quad (O_F^\dagger)_H^{(S)}=\left\{(O_F)_H^{(S)}\right\}^\dagger\text{,}
\end{equation}
\begin{equation}
\vert\psi)_H^{(S)}=U_H^{(S)}\vert\psi\rangle;\quad {}^{(S)}_{\quad H}(\psi\vert=\left\{\vert\psi)_H^{(S)}\right\}^\dagger\text{.}
\end{equation}
\end{subequations}
However, even if these are inverse-transformed, they do not revert back to the original:
\begin{equation}
\left\{U_H^{(S)}\right\}^\dagger (O_F)_H^{(S)}U_H^{(S)}=\hat T_F^{(S)}O_F\hat T_F^{(S)}\text{,}\quad \left\{U_H^{(S)}\right\}^\dagger\vert\psi )_H^{(S)}=\hat T_F^{(S)}\vert\psi\rangle\text{.}
\end{equation}
To obtain a one-to-one correspondence, both the operators and state vectors must be limited to the fermion subspace projected by $\hat T_F^{(S)}$. However, adding only this restriction is not sufficient to obtain an isomorphic mapping, because
\begin{equation}
\begin{array}{lll}
\hat T_F^{(S)}O_FO'_F\hat T_F^{(S)}&=&\hat T_F^{(S)}O_F\hat T_F^{(S)} O'_F\hat T_F^{(S)}+\hat T_F^{(S)}O_F(\hat 1_F-\hat T_F^{(S)})O'_F\hat T_F^{(S)}
\\
&\neq &\hat T_F^{(S)}O_F\hat T_F^{(S)} O'_F\hat T_F^{(S)}\text{.}
\end{array}
\end{equation}
Therefore, to obtain an isomorphic mapping, the quasiparticle pair operators themselves must be restricted by $\hat T_F^{(S)}$ before their product is taken. In this way, we obtain the isomorphic mapping:
\begin{equation}
\begin{array}{lll}
(\hat T_F^{(S)}O_F\hat T_F^{(S)}\hat T_F^{(S)}O'_F\hat T_F^{(S)})_H^{(S)}&=&(\hat T_F^{(S)}O_F\hat T_F^{(S)}O'_F\hat T_F^{(S)})_H^{(S)}
\\
&=&(\hat T_F^{(S)}O_F\hat T_F^{(S)})_H^{(S)}(\hat T_F^{(S)}O'_F\hat T_F^{(S)})_H^{(S)}\text{.}
\end{array}
\end{equation}
Through this mapping, for state vectors belonging to the subspaces projected by $\hat T_B^{(S)}$ and $\hat T_F^{(S)}$, we obtain
\begin{equation}
\begin{array}{lll}
{}^{(S)}_{\quad H}(\psi'\vert (O_F)_H^{(S)}\vert\psi)_H^{(S)}=\langle\psi'\vert O_F\vert\psi\rangle\text{,}
\\
{}^{(S)}_{\quad H}(\psi'\vert (O_F)_H ^{(S)}(O'_F)_H^{(S)}\vert\psi)_H^{(S)}=\langle\psi'\vert O_F\hat T_F^{(S)}O'_F\vert\psi\rangle\text{.}
\end{array}
\end{equation}

For the non-Hermitian-type mapping, the following mapping operators are adopted:
\begin{equation}
U_1^{(S)}=\sum_{N=0}^\infty\sum_{a\neq a_0}z_a(N)^{\frac 12}\vert N; a)\langle N; a\vert,\quad U_2^{(S)}=\sum_{N=0}^\infty\sum_{a\neq a_0}z_a(N)^{-\frac 12}\vert N; a)\langle N; a\vert\text{.}
\end{equation}
These satisfy the relations:
\begin{equation}
U_1^{(S)}\left\{U_2^{(S)}\right\}^\dagger =\hat T_B^{(S)},\quad
\left\{U_2^{(S)}\right\}^\dagger U_1^{(S)}=\hat T_F^{(S)}\text{.}
\end{equation}
The following relations, corresponding to the Hermitian type, hold:
\begin{subequations}
\begin{equation}
(O_F)_{NH}^{(S)}=U_1^{(S)}O_F\left\{U_2^{(S)}\right\}^\dagger;\quad (O_F^\dagger)_{NH}^{(S)}\neq\left\{(O_F)_{NH}^{(S)}\right\}^\dagger\text{,}
\end{equation}
\begin{equation}
\vert\psi)_R^{(S)}=U_1^{(S)}\vert\psi\rangle, {}^{(S)}_{\quad L}(\psi\vert =\rangle\psi\vert\left\{U_2^{(S)}\right\}^\dagger ;\quad {}^{(S)}_{\quad L}(\psi\vert=\left\{\vert\psi)_R^{(S)}\right\}^\dagger\text{.}
\end{equation}
\end{subequations}
The conditions for this transformation to be one-to-one and an isomorphic mapping are the same as for the Hermitian type. Finally, through this mapping, for state vectors belonging to the subspaces projected by $\hat T_B^{(S)}$ and $\hat T_F^{(S)}$, we obtain:
\begin{equation}
\begin{array}{lll}
{}^{(S)}_{\quad L }(\psi'\vert (O_F)_{NH}^{(S)}\vert\psi)_R^{(S)}=\langle\psi'\vert O_F\vert\psi\rangle\text{,}
\\
{}^{(S)}_{\quad L}(\psi'\vert (O_F)_{NH}^{(S)}(O'_F)_D^{(S)}\vert\psi)_R^{(S)}=\langle\psi'\vert O_F\hat T_F^{(S)}O'_F\vert\psi\rangle\text{.}
\end{array}
\end{equation}

One of the advantages of mapping only the fermion subspace is that a boson state vector that does not include the effects of the Pauli exclusion principle at all becomes a physical state vector by appropriately limiting the number and types of phonon excitations. That is, by defining
\begin{equation}
\breve 1_B=\sum_{N=0}^{N_{max}}\sum_t\vert N; t)(N; t\vert\text{,}
\end{equation}
the following relations hold by appropriately choosing $\{t\}$ and $N_{max}$:
\begin{equation}
\breve 1_B\hat T_B^{(S)}=\hat T_B^{(S)}\breve 1_B=\breve 1_B\hat T_B^{(S)}\breve 1_B=\breve 1_B\text{.}
\end{equation}
This ensures that only the mapped operators bear the effect of the Pauli exclusion principle.

Another advantage lies in the convergence of the boson expansion.
Appropriate selection of the phonon excitation modes makes it possible to choose the part where the effect of the Pauli exclusion principle is weak.
The mapped phonon operators allow for the stepwise inclusion of the Pauli exclusion principle effects as a small parameter expansion, with the zeroth-order approximation taken as bosons. This is not the case, however, if the number of phonon excitations increases too much. This point was not considered in conventional boson expansion theories and was explicitly handled for the first time by the norm operator method.

The mapping operators have been obtained by Kishimoto and Tamura (KT-3) for the Hermitian type, and by Takada for the non-Hermitian type:
\begin{subequations}
\begin{equation}
(X_t')_H^{(S)}=b_t^\dagger -\frac 14\sum_{t_1t'_1t'_2}Y(t't'_1t'_2t_1)b_{t_1}^\dagger b_{t'_1}b_{t'_2}\text{,}
\end{equation}
\begin{equation}
(B_q)_H^{(S)}=\sum_t\sum_{t'}\Gamma_q^{t't}b_t^\dagger b_{t'}\text{,}
\end{equation}
\end{subequations}
\begin{subequations}
\begin{equation}
(X_{t'})_{NH}^{(S)}=(X_{t'})_D\hat T_B^{(S)}\text{,}
\end{equation}
\begin{equation}
(X_t^\dagger)_{NH}^{(S)}=(X_t^\dagger)_D\hat T_B^{(S)}\text{,}
\end{equation}
\begin{equation}
(B_q)_{NH}^{(S)}=(B_q)_D\hat T_B^{(S)}\text{,}
\end{equation}
\end{subequations}
Here,
\begin{subequations}
\label{eq:dbexp}
\begin{equation}
(X_{t'})_D^{(S)}=b_{t'}\text{,}
\end{equation}
\begin{equation}
\label{eq:xdaggerd}
(X_t^\dagger)_D^{(S)}=b_t^\dagger-\frac 12\sum_{t_1t_2}\sum_{t'_1}Y(tt_1t_2t'_1)b_{t_1}^\dagger b_{t_2}^\dagger b_{t'_1}\text{,}
\end{equation}
\begin{equation}
\label{eq:bqd}
(B_q)_D^{(S)}=\sum_t\sum_{t'}\Gamma_q^{t't}b_t^\dagger b_{t'}\text{.}
\end{equation}
\end{subequations}

The former is obtained as an infinite expansion, while the latter is a finite expansion.
The projection operator $\hat T_B^{(S)}$ onto the physical space appears in Takada's mapping for each operator, but not in Kishimoto and Tamura's (KT-3). This is because KT-3 presupposed that the norm matrix eigenvalues of multi-phonon state vectors would not become zero, a condition which would eventually include zero eigenvalues as the number of phonon excitations increased beyond a certain point. This premise does not actually hold in practice. It is equivalent to assuming that
\begin{equation}
\hat T_B^{(S)}=\hat 1_B^{(S)}\text{,}\quad \hat 1_B^{(S)}=\sum_{N=0}^\infty\sum_t\vert N; t)(N; t\vert
\end{equation}
holds true.
Furthermore, both expansions are derived by ignoring the excitation modes $\{\bar t\}$ (NAMD) other than the phonon excitation modes $\{t\}$ adopted for the construction of the subspace. Takada derives the finite expansion by comprehensively adopting NAMD. In contrast, Kishimoto and Tamura calculate the expansion by partially rejecting it initially, demonstrating that the terms ignored by NAMD appear in the second and subsequent terms of the infinite expansion of $(B_q)_H^{(S)}$, and then subsequently adopt NAMD.
If NAMD is adopted, the second and subsequent terms vanish, and the boson expansion of the scattering operator becomes a finite expansion, which matches the Dyson boson expansion obtained by Takada.
In KT-3 as well, NAMD is regarded as a "good approximation." However, although there is an explanation attempting to support this, it is not persuasive and does not provide proof guaranteeing the validity of the approximation.

In Takada's method, NAMD is performed using a method called the "closed algebra approximation." This imposes the following conditions to exclude the non-adopted phonon excitation modes $\{\bar t\}$ as boson excitations:
\begin{subequations}
\begin{equation}
\label{eq:caa1}
a_\alpha^\dagger a_\beta^\dagger =\sum_\mu\psi_\mu(\alpha\beta)X_\mu\approx \sum_t\psi_t(\alpha\beta)X_t\text{,}
\end{equation}
\begin{equation}
\label{eq:caa2}
[[X_{t_1}, X_{t_2}^\dagger], X_{t_3}^\dagger ]=-\sum_\mu Y(t_1t_2t_3\mu)X_\mu^\dagger\approx -\sum_t Y(t_1t_2t_3t)X_t^\dagger\text{.}
\end{equation}
\end{subequations}
For Eq. (\ref{eq:caa2}) to hold strictly, $Y(t_1t_2t_3\bar t)=0$ must be satisfied for all $\{t\}$ and $\{\bar t\}$.
Furthermore, when applying the closed algebra approximation to
\begin{equation}
[[a_\beta a_\alpha, X_{t_1}^\dagger], X_{t_2}^\dagger]
=-\sum_\mu\psi_\mu(\beta\alpha)[[X_\mu, X_{t_1}^\dagger], X_{t_2}^\dagger]]
=-\sum_{\mu\mu'}Y(\mu t_1t_2\mu')X_{\mu'}^\dagger\text{,}
\end{equation}
we obtain
\begin{equation}
[[a_\beta a_\alpha, X_{t_1}^\dagger], X_{t_2}^\dagger] \approx -\sum_t\psi_t(\beta\alpha)\sum_{t'}Y(tt_1t_2t')X_{t'}^\dagger\text{.}
\end{equation}
A comparison before and after the application shows that for the closed algebra approximation to hold strictly, $Y(\bar tt_1t_2\bar t')=0$ must be satisfied for all $\{t\}$ and $\{\bar t\}$.
Conversely, if $Y(t_1t_2t_3\bar t)=0$ and $Y(\bar tt_1t_2\bar t')=0$ hold for all $\{t\}$ and $\{\bar t\}$, the closed algebra approximation holds strictly. Considering the symmetry of $Y(\mu_1\mu_2\mu_3\mu_4)$ with respect to its arguments in Eq. (\ref{eq:Ysym}), the strict validity of the closed algebra approximation means that all parts involving $\{\bar t\}$ in Eq. (\ref{eq:doublecom}) are ignored. This implies a complete application of NAMD when obtaining the boson expansion.
Additionally, from
\begin{equation}
\langle\mu'_1\mu'_2\vert\mu_1\mu_2\rangle=(\mu'_1\mu'_2\vert\mu_1\mu_2)-{\mathcal N}(2; \mu')^{-\frac 12}Y(\mu_1\mu'_1\mu'_2\mu_2){\mathcal N}(2; \mu)^{-\frac 12}\text{,}
\end{equation}
it can be said that the necessary and sufficient conditions for the closed algebra approximation to hold strictly are $\langle t'\bar t'\vert t_1t_2\rangle=0$ and $\langle\bar t'_1\bar t'_2\vert t_1t_2\rangle=0$.
These conditions play a crucial role in the following norm operator method.

\subsection{Relationship between Whole Space Mapping and Subspace Mapping}

Conventional boson expansion theories discuss the relationship between whole space mapping and subspace mapping, but their treatment is insufficient and causes confusion \cite{LH75, Ta01}.

To relate WFSM and FSSM, we rewrite the boson creation and annihilation operators of the WFSM as follows, mirroring how the phonon operators are constructed from the quasiparticle pair operators:
\begin{equation}
\label{eq:babbmu}
b_\mu=\sum_{\alpha<\beta}\psi_\mu(\alpha\beta)b_{\alpha\beta}\text{,} \quad
b_\mu^\dagger=\sum_{\alpha<\beta}\psi_\mu(\alpha\beta)b_{\alpha\beta}^\dagger\text{.}
\end{equation}
WFSM adopts all these bosons corresponding to all quasiparticle pair excitation modes. In contrast, FSSM adopts only $b_t$ and $b_t^\dagger$ among these. That is, FSSM introduces only those bosons whose excitation modes correspond to the phonon excitation modes adopted to construct the multi-phonon state vectors that form the basis of the mapped subspace.

Therefore, the problem should be how to handle the boson excitations $\{\bar t\}$ not introduced in the latter mapping (FSSM) among those introduced in the former mapping (WFSM), when deriving the results of the FSSM from those of the WFSM. However, conventional methods have completely ignored this point, and the only means to obtain the FSSM from the WFSM has been to simply limit the boson operators to $b_{t'}$ and $b_t^\dagger$ adopted in the FSSM and discard the rest.
It has been argued that the mapping operator of the FSSM cannot be obtained from the results of the WFSM. Specifically, it is claimed that there is only one operator,
\begin{equation}
\breve 1_B^{(S)}=\sum_{N=0}^\infty\vert N; t)(N; t\vert\text{,}
\end{equation}
that obtains the FSSM from the WFSM, and applying it to the mapping operator of the WFSM results in
\begin{equation}
\breve 1_B^{(S)}\hat U_H^{(W)}\neq U_H^{(S)}\text{.}
\end{equation}
If this claim holds true, it is clear that it similarly holds for the mapping operators $U_1^{(W)}$ and $U_2^{(W)}$ of the Dyson boson expansion method.
As a result of this, the following results are derived for the mapping of operators:
\begin{equation}
\breve 1_B^{(S)}(O_F)_H^{(W)}\breve 1_B^{(S)}\neq (O_F)_H^{(S)}\text{,} \quad \breve 1_B^{(S)}(O_F)_{NH}^{(W))}\breve 1_B^{(S)}\neq (O_F)_{NH}^{(S)}\text{.}
\end{equation}
Nevertheless, in the Dyson boson expansion method, instead of addressing $(O_F)_{NH}^{(W)}=(O_F)_D{(W)}\hat T_B^{(W)}$ itself, attention is drawn to $(O_F)_D^{(W)}$, and by noting that
\begin{equation}
\breve 1_B^{(S)}(O_F)_D^{(W)}\breve 1_B^{(S)}=(O_F)_D^{(S)}
\end{equation}
holds true, the consistency of the algebraic structure between the results of the WFSM (limited as above) and the FSSM is emphasized, and the effectiveness is advocated by carrying over the name "skeleton boson realizations \cite{KV88}."
On the other hand, regarding the physical subspace, it is argued that the Park operator \cite{P87}, which was introduced in the WFSM to determine whether a boson state vector is physical, is not effective in the FSSM.
The reason for this is discussed by attributing it to the difference between the projection operators onto the physical subspace, $\hat T_B^{(W)}$ for the WFSM and $\hat T_B^{(S)}$ for the FSSM. They ultimately justify the results obtained by the FSSM and defend the effectiveness of the skeleton boson realization, but the argument is not clear.

These inadequacies and confusions cannot be resolved by conventional boson expansion theories that adopt NAMD.
In the next section, we analyze this confused situation using the norm operator method  \cite{NR23}, which does not presuppose NAMD. Based on this analysis, we present clear conclusions regarding this problem.

\section{Analysis and Conclusion by the Norm Operator Method}
\label{normac}

The norm operator method is a method for obtaining concrete boson expansions without presupposing NAMD, by utilizing a norm operator that includes the multi-phonon norm matrix, which explicitly reflects the effects of the Pauli exclusion principle.

This method also adopts a mapping for the fermion subspace mapping (FSSM) that not only limits the types of phonon excitation modes, as in conventional methods, but also simultaneously limits the number of phonon excitations, which had not been attempted previously. As a result, it has become possible to map only the fermion subspace where the ideal boson state vectors, which contain no effects of the Pauli exclusion principle, become the physical state vectors.

In that mapping, we established a method to obtain the boson expansion theory (BET) more easily compared to conventional methods, and we can now determine terms up to higher orders that were previously unobtainable.

The introduction of the norm operator also clarified the role played by NAMD through analysis, leading to results that overturn conventional wisdom.

Specifically, by clarifying the relationship between the norm operator that incorporates all phonon excitation modes (limiting only the number of phonon/boson excitations) and the norm operator that also limits the phonon excitation modes, we established a procedure for obtaining the small parameter expansion. We showed that when the small parameter expansion holds, all expansions, regardless of whether they are Hermitian or non-Hermitian type, are given as infinite expansions. On the other hand, we clarified that when NAMD holds strictly, the small parameter expansion does not hold, and the expansion becomes a substantially finite expansion for all mappings, including the Hermitian type, not limited to the non-Hermitian type mapping that gives the Dyson boson expansion.

In this section, we analyze the whole fermion space mapping (WFSM) and the fermion subspace mapping (FSSM) using this norm operator method and derive the correct conclusions.

\subsection{The Norm Operator Method}
The mapping operator of the norm operator method is given as follows:
\begin{equation}
\label{eq:bmopzubar}
U_\xi=\hat Z^{\xi-\frac 12}\widetilde U\text{,}
\end{equation}
where $\hat Z$ is the norm operator, and its definition is as follows:
\begin{subequations}
\begin{equation}
\label{eq:normop}
\hat Z=\sum_{N=0}^{N_{max}}\hat Z(N)\text{,}
\end{equation}
\begin{equation}
\label{eq:normop2}
\begin{array}{lll}
\hat{Z}(N)&=&\displaystyle\sum_{t t'}\vert N, t)\langle N;
t\vert N; t'\rangle ( N; t' \vert
\\
&=&
\displaystyle\sum_{t_1\leq\cdots \leq t_N}\sum_{t'_1\leq\cdots
\leq t'_N}\vert t_1 \cdots t_N)\langle t_1\cdots t_N\vert
t'_1\cdots t_N\rangle (t'_1\cdots t'_N \vert\text{.}
\end{array}
\end{equation}
\end{subequations}
Also, $\vert N; a)$ and $z_a(N)$ are the eigenvectors and their eigenvalues of the norm operator $\hat Z$, respectively. Using these, $\hat Z^\eta$ is defined as follows:
\begin{equation}
\hat Z^\eta=\sum_{N=0}^{N_{max}}\hat Z(N)^\eta;\quad \hat Z(N)^\eta=\sum_{a\neq a_0}\vert N; a)z_a(N)^\eta (N; a\vert\text{.}
\end{equation}
Furthermore, the definition of $\widetilde U$ is as follows:
\begin{subequations}
\label{eq:tildempop}
\begin{equation}
\label{eq:tildempop1}
\widetilde {U}=\sum_{N=0}^{N_{max}}{\widetilde U}(N)\text{,}
\end{equation}
\begin{equation}
\label{eq:tildempop2}
\begin{array}{lll}
{\widetilde U}(N)&=&\displaystyle\sum_{t}\vert N; t)\langle N; t\vert
\\
&=&
\displaystyle\sum_{t_1\leq t_2\leq\cdots\leq t_N}\vert t_1 t_2\cdots
t_N)\langle t_1 t_2\cdots t_N\vert\text{.}
\end{array}
\end{equation}
\end{subequations}
The following relations are satisfied:
\begin{equation}
\label{eq:utftbh}
U_{-\xi}^\dagger U_{\xi}=\hat T_F,\qquad U_{\xi}U_{-\xi}^\dagger =\hat T_B\text{,}
\end{equation}
where
\begin{equation}
\label{eq:unitf}
\hat T_F=\sum_{N=0}^{N_{max}}\hat T_F(N);\qquad
\hat T_F(N)=\displaystyle\sum_{a\neq a_0}\vert N; a\rangle\langle N; a\vert\text{,}
\end{equation}
\begin{equation}
\hat T_B=\sum_{N=0}^{N_{max}}\hat T_B(N);\quad\hat T_B(N)=\sum_{a\neq a_0}\vert N; a)(N; a\vert\text{.}
\end{equation}
In addition, we define the following operator:
\begin{equation}
\label{eq:breve1B}
\breve 1_B=\sum_{N=0}^{N_{max}}\hat 1_B(N);\qquad \hat 1_B(N)=\sum_t\vert N; t)(N;t\vert\text{.}
\end{equation}
If $\hat Z(N)$ has even one zero eigenvalue, then $\hat T_B(N)\neq \hat 1_B(N)$ and hence $\hat T_B\neq \breve 1_B$. Otherwise, they match one another.

The state vectors and operators of fermion space are mapped onto those of boson subspace as follows:
\begin{subequations}
\label{eq:ximap}
\begin{equation}
\label{eq:ximap1}
\vert \psi')_{\xi}= U_{\xi}\vert\psi'\rangle,\qquad {}_{-\xi} (\psi\vert =\langle\psi\vert U_{-\xi}^\dagger\text{,}
\end{equation}
\begin{equation}
\label{eq:ximap2}
(O_F)_{\xi}=U_{\xi}O_FU_{-\xi}^\dagger\text{.}
\end{equation}
\end{subequations}
These satisfy the following relations:
\begin{subequations}
\label{eq:ximappm}
\begin{equation}
\label{eq:ximappm1}
\vert \psi')_{\xi}=\left\{{}_\xi(\psi'\vert\right\}^\dagger,\qquad {}_{-\xi} (\psi\vert =\left\{\vert\psi)_{-\xi}\right\}^\dagger\text{,}
\end{equation}
\begin{equation}
\label{eq:ximappm2}
(O_F)_{-\xi}=\left\{(O_F^\dagger)_{\xi}\right\}^\dagger\text{.}
\end{equation}
\end{subequations}
The mapping is of the Hermitian type when $\xi=0$ and, in other cases, of the non-Hermitian type. 
For the state vectors, $\vert\psi\rangle$ and $\vert\psi'\rangle$, which belong to the fermion subspace projected by $\hat T_F$,
\begin{equation}
\label{eq:mteqh}
\begin{array}{lll}
\langle\psi\vert O_F\vert\psi'\rangle
&=&\langle\psi\vert\hat T_F O_F\hat T_F\vert\psi'\rangle
\\
&=&\langle\psi\vert U_{-\xi}^\dagger U_\xi O_FU_{-\xi}^\dagger U\vert\psi'\rangle
\\
&=&{}_{-\xi}(\psi\vert(O_F)_\xi\vert\psi')_\xi\text{.}
\end{array}
\end{equation}
That is, the matrix element of the fermion subspace becomes equal to that of the corresponding boson subspace.

A one-to-one correspondence exists between the fermion subspace projected by $\hat T_F$ and the boson subspace by $\hat T_B$.

For this to become an isomorphic mapping, the phonon and scattering operators themselves must be restricted by $\hat T_F$ before their product is taken, similar to the conventional boson expansion theory. As a result, we obtain the following relationship:
\begin{equation}
(\hat T_FO_F\hat T_F\hat T_FO'_F\hat T_F)_\xi=(\hat T_FO_F\hat T_FO'_F\hat T_F)_\xi=(\hat T_FO_F\hat T_F)_\xi(\hat T_FO'_F\hat T_F)_\xi\text{.}
\end{equation}
Through this mapping, for state vectors belonging to the subspaces projected by $\hat T_B$ and $\hat T_F$, we obtain the following relations:
\begin{equation}
\begin{array}{lll}
{}_{-\xi}(\psi'\vert (O_F)_\xi\vert\psi)_\xi=\langle\psi'\vert O_F\vert\psi\rangle\text{,}
\\
{}_{-\xi}(\psi'\vert (O_F)_\xi(O'_F)_\xi\vert\psi)_\xi=\langle\psi'\vert O_F\hat T_FO'_F\vert\psi\rangle\text{.}
\end{array}
\end{equation}

The norm operator method becomes a practical boson expansion method by appropriately limiting the types and number of phonon excitation modes. Furthermore, conventional boson expansion theories can be derived by removing these limitations. Specifically, by removing only the limitation on the number of phonon excitations, the conventional FSSM is obtained. If we additionally remove the limitation on the types of phonon excitation modes and adopt all types, WFSM is obtained.

We denote $U_\xi^{(A)}$, $\hat Z^{(A)}$, and ${\widetilde U}^{(A)}$ as the operators obtained by keeping the limitation on the number of phonon (= boson) excitations but removing the limitation on the types of phonon excitation modes to adopt all modes. Furthermore, the WFSM obtained by removing the limitation on the number of phonon excitations is denoted as $U_\xi^{(W)}$, $\hat Z^{(W)}$, and ${\widetilde U}^{(W)}$.

To obtain a practical boson expansion method, focusing solely on the norm operator $\hat Z$ is insufficient. $\hat Z$ is one component of $\hat Z^{(A)}$, which consists of all excitation modes, and the structure of $\hat Z^{(A)}$ determines the structure of $\hat Z$.
The relation between $\hat Z^{(A)}$ and $\hat Z$ is given by:
\begin{subequations}
\label{eq:zazwzd}
\begin{equation}
\hat Z^{(A)}=\hat Z+\hat W+\hat W^\dagger+\hat Z'\text{,}
\end{equation}
where
\begin{equation}
\hat Z=\breve 1_B\hat Z^{(A)}\breve 1_B\text{,}
\quad\hat W=\breve 1_B\hat Z^{(A)}(\breve 1_B^{(A)}-\breve 1_B)\text{,}
\quad\hat Z'=(\breve 1_B^{(A)}-\breve 1_B)\hat Z^{(A)}(\breve 1_B^{(A)}-\breve 1_B)\text{.}
\end{equation}
\end{subequations}
$\hat Z^{(A)}$ is expressed as follows:
\begin{equation}
\label{eq:normopA}
\hat Z^{(A)}=(2\hat N_B^{(A)}-1)!!\hat T_B^{(A)};\quad\hat N_B^{(A)}=\sum_\mu b_\mu^\dagger b_\mu\text{,}
\end{equation}
Under this condition, the structure of $\hat Z$ is determined by the behavior of $\hat W$ and $\hat Z'$.

For the small parameter expansion to hold, where the phonons become bosons in the zeroth approximation, the norm operator must become a unit operator; that is, $\hat 1_B$ for the whole fermion space mapping (WFSM) and $\breve 1_B$ for the fermion subspace mapping (FSSM). Therefore, it can be seen from Eq. (\ref{eq:normopA}) that the small parameter expansion does not hold for the WFSM.
In the case of the FSSM, on the other hand, the condition is realized by appropriately limiting the type and number of phonon excitation modes, and a small parameter expansion becomes possible. The specific method for this must be determined under the condition of Eq. (\ref{eq:normopA}), and that has been achieved. It is noteworthy that when the small parameter expansion holds, it is shown that any type of boson expansion is obtained as an infinite expansion, regardless of whether it is Hermitian or non-Hermitian. The contributions of phonon excitations ignored by NAMD are then renormalized into the coefficients of the boson expansion.

Utilizing the relationship between $\hat Z^{(A)}$ and $\hat Z$ allows us to draw clear conclusions about cases where NAMD, treated as a "good approximation" without clear justification in conventional boson expansion theories, and the equivalent closed algebra approximation, hold strictly. NAMD, the closed algebra approximation, and $\langle t'\bar t'\vert t_1t_2\rangle=0$, $\langle\bar t'_1\bar t'_2\vert t_1t_2\rangle=0$ are necessary and sufficient conditions for each other to hold.

If $\langle t'\bar t'\vert t_1t_2\rangle=0$ and $\langle\bar t'_1\bar t'_2\vert t_1t_2\rangle=0$ holds, then $\hat W=0$ is satisfied. That is, if NAMD or the equivalent closed algebra approximation holds strictly, we must have $\hat W=0$.
And if $\hat W=0$ holds, then
\begin{subequations}
\begin{equation}
\hat Z=\breve 1_B\hat Z^{(A)}=\hat Z^{(A)}\breve 1_B=\breve 1_B\hat Z^{(A)}\breve 1_B\text{,}
\end{equation}
holds, and we obtain
\begin{equation}
\hat T_B=\breve 1_B\hat T_B^{(A)}=\hat T_B^{(A)}\breve 1_B=\breve 1_B \hat T_B^{(A)}\breve 1_B\text{.}
\end{equation}
\end{subequations}
Finally,
\begin{equation}
\label{eq:znnons2}
\hat Z=(2\hat N_B-1)!!\hat T_B;\quad\hat N_B =\sum_t b_t^\dagger b_t
\end{equation}
is derived.
This shows that the small parameter expansion where $\breve 1_B$ is the zeroth approximation for $\hat Z$ is impossible.

When NAMD holds, the mapping operator is as follows:
\begin{equation}
\label{eq:ofxi1}
(O_F)_\xi=\hat Z^{\xi-\frac 12}(O_F)_D\hat Z^{-\xi+\frac 12}\text{,}
\end{equation}
where $(O_F)_D$ is the Dyson boson expansion. These expansions satisfy $(O_F)_D=(O_F)_D^{(S)}$ and are the same regardless of the limitation on the number of phonon excitations.
Furthermore, the following relationship holds:
\begin{subequations}
\label{eq:ofofd}
\begin{equation}
\label{eq:oftfofd}
O_FO'_F=O_F\hat T_FO'_F\text{.}
\end{equation}
Therefore, the following relation holds:
\begin{equation}
\label{eq:ofxi2ofof}
(O_FO'_F)_\xi=(O_F)_\xi(O'_F)_\xi\text{.}
\end{equation}
\end{subequations}
Thus, if NAMD holds, the mapping becomes an isomorphic mapping even without imposing restrictions beforehand, such as $\hat T_BO_F\hat T_B$ and $\hat T_BO'_F\hat T_B$, on the phonon and scattering operators.

\subsection{Derivation of Subspace Mapping from Whole Space Mapping}

The difference between the whole fermion space mapping (WFSM) and the fermion subspace mapping (FSSM) lies in the boson excitations they handle. The former introduces bosons corresponding to all quasiparticle pair excitations, whereas the latter introduces only a part of them. Therefore, to derive the FSSM from the WFSM, the WFSM boson excitations that are not introduced in the FSSM must be treated correctly.

This was impossible with conventional boson expansion theories that adopted NAMD. However, with the norm operator method, which correctly incorporates phonon excitations overlooked by NAMD, this is realized simply and clearly by rewriting the following mapping operator:
\begin{equation}
U_\xi=U_\xi\hat 1_F=U_\xi\left\{U_{-\eta}^{(W)}\right\}^\dagger U_\eta^{(W)}=\hat P_{(\xi;\eta)} U_\eta^{(W)}\text{.}
\end{equation}
From this, the following relationship is obtained:
\begin{equation}
(O_F)_\xi=\hat P_{(\xi;\eta)}(O_F)_\eta^{(W)}\hat P_{(-\xi;-\eta)}^\dagger\text{.}
\end{equation}
The operator $\hat P_{(\xi;\eta)}$ can be rewritten as follows:
\begin{equation}
\begin{array}{lll}
\hat P_{(\xi;\eta)}&=&U_\xi\left\{U_{-\eta}^{(W)}\right\}^\dagger 
\\
&=&U_\xi\left\{U_{-\eta}^{(A)}\right\}^\dagger 
\\
&=&\hat Z^{\xi-\frac 12}{\widetilde U}\left\{{\widetilde U}^{(A)}\right\}^\dagger\left\{\hat Z^{(A)}\right\}^{-\eta-\frac 12}\text{.}
\end{array}
\end{equation}

This relationship provides a new perspective to the norm operator method. The physical subspace obtained by the whole fermion space mapping (WFSM) is a complete replica of the entire fermion space. The above relationship shows that the physical subspace obtained by the fermion subspace mapping (FSSM) can be acquired if the contributions of phonons, which are discarded by NAMD, are correctly renormalized from the WFSM physical subspace using $\hat P_{(\xi;\eta)}$. The contributions of these renormalized bosons are then reflected in the coefficients of the boson expansion.

The norm operator method also provides a clear explanation for NAMD. The following relations hold:
\begin{subequations}
\begin{equation}
{\widetilde U}\left\{{\widetilde U}^{(A)}\right\}^\dagger={\widetilde U}{\widetilde U}^\dagger =\hat Z\text{,}
\end{equation}
\begin{equation}
\left\{\hat Z^{(A)}\right\}^{-\eta-\frac 12}=\left\{\hat Z+\hat Z'\right\}^{-\eta-\frac 12}=\hat Z^{-\eta-\frac 12}+\hat Z'^{-\eta-\frac 12}\text{,}
\end{equation}
\end{subequations}
and therefore, the following relationship holds:
\begin{equation}
\hat P_{(\xi;\eta)}=\hat Z^{\xi-\eta}\text{.}
\end{equation}
Furthermore, if we take $\xi=\eta$, then $\hat P_{(\xi;\xi)}=\hat T_B$ is obtained. If we then appropriately limit the types and number of excitation modes, we obtain $\hat P_{(\xi;\xi)}=\breve 1_B$.
Therefore, if NAMD holds, the mapping operator for the fermion subspace mapping (FSSM) can be obtained simply by limiting the boson excitations of the whole fermion space mapping (WFSM) operator. This conclusion also holds for conventional FSSM where no limitation is imposed on the number of phonon excitations. Conventional boson expansion theories adopt NAMD. Nevertheless, they did not reach this conclusion because the influence of NAMD was not correctly understood.

On the other hand, if NAMD holds, all types of mappings become isomorphic mappings, not limited to the mapping adopted in the Dyson boson expansion method. Therefore, the concept of "skeleton boson realizations" holds true for all types of boson expansion methods, not just limited to the Dyson boson expansion method, provided NAMD is satisfied.

Furthermore, if NAMD holds, the Park operator is effective for the fermion subspace mapping (FSSM) as well as for the whole fermion space mapping (WFSM). In general, $\hat T_B\neq \breve 1_B\hat T_B^{(W)}$ holds, but if NAMD holds, $\hat T_B=\breve 1_B\hat T_B^{(W)}$ is satisfied. As a result, the following equation holds:
\begin{equation}
\begin{array}{lll}
\hat S\hat T_B&=&\hat S\breve 1_B\hat T_B^{(W)}=\breve 1_B\hat S\hat T_B^{(W)}
\\
&=&\breve 1_B(\hat N_F^2-\hat N_F-\sum_{\alpha\beta}a_\alpha^\dagger a_\beta^\dagger a_\alpha a_\beta)_D^{(W)}\hat T_B^{(W)}=0\text{.}
\end{array}
\end{equation}
Here, in deriving $\hat T_B=\breve 1_B\hat T_B^{(W)}$, we utilized the following relations, which hold under the premise of $\hat W=0$:
\begin{subequations}
\begin{equation}
\breve 1_B\hat T_B^{(W)}=\breve 1_B\lim_{N_{max}\rightarrow\infty}\hat T_B^{(A)}=\breve 1_BU_\xi^{(A)}\left\{U_{-\xi}^{(A)}\right\}^\dagger\text{,}
\end{equation}
\begin{equation}
\breve 1_BU_\xi^{(A)}=\breve 1_B\left\{\hat Z^{(A)}\right\}^{\xi-\frac 12}\left\{{\widetilde U}^{(A)}\right\}^\dagger
=\hat Z^{\xi-\frac 12}\breve 1_B{\widetilde U}^{(A)}=U_\xi\text{,}
\end{equation}
\begin{equation}
U_\xi\left\{{\widetilde U}^{(A)}\right\}^\dagger=U_\xi{U_{-\xi}}^\dagger\text{.}
\end{equation}
\end{subequations}
Since the Dyson boson expansion presupposes the validity of NAMD, the assertion that the Park operator is not effective in the FSSM is incorrect.

In the fermion subspace mapping (FSSM), if the types and number of phonon excitation modes are chosen appropriately, $\hat T_B=\breve 1_B$ holds, and the ideal boson state vectors $\vert N; t)$ become physical state vectors. In the general case, however, the Park operator does not identify these as physical states. This becomes clear if the Park operator is rewritten using Eq. (\ref{eq:babbmu}) as follows:
\begin{equation}
\hat S=4\sum_{\mu_1\mu_2}b_{\mu_1}^\dagger b_{\mu_2}^\dagger b_{\mu_1}b_{\mu_2}+\sum_{\mu_1\mu_2\mu'_1\mu'_2}Y(\mu'_2\mu_1\mu_2\mu'_1)b_{\mu_1}^\dagger b_{\mu_2}^\dagger b_{\mu'_1}b_{\mu'_2}\text{.}
\end{equation}
However, when NAMD holds, it is also shown to be correctly regarded as a physical state vector by the following rewriting: $Y(\mu'_2\mu_1\mu_2\mu'_1)$ satisfies the following relation,
\begin{equation}
\langle\langle\mu'_1\mu'_2\vert\mu_1\mu_2\rangle\rangle
=((\mu'_1\mu'_2\vert\mu_1\mu_2))-Y(\mu'_2\mu_1\mu_2\mu'_1)\text{,}
\end{equation}
where $\vert \mu_1\mu_2\rangle\rangle={\mathcal N}_B^\frac 12(\mu_1,\mu_2)\vert \mu_1\mu_2\rangle$ and $\vert \mu_1\mu_2))={\mathcal N}_B^\frac 12(\mu_1,\mu_2)\vert \mu_1\mu_2)$.
And it can be rewritten further as:
\begin{equation}
\langle\langle\mu'_1\mu'_2\vert\mu_1\mu_2\rangle\rangle
=((\mu'_1\mu'_2\vert\hat Z^{(A)}\vert\mu_1\mu_2))\text{.}
\end{equation}
If NAMD holds, we obtain the following equation by setting $\hat W=0$ in Eq. (\ref{eq:zazwzd}) and substituting Eq. (\ref{eq:normopA}):
\begin{equation}
\hat Z^{(A)}=\hat Z+\hat Z'=(2\hat N_B-1)!!\hat T_B+\hat Z'\text{.}
\end{equation}
Using these relations, the Park operator is expressed as follows:
\begin{equation}
\label{eq:sanamd}
\begin{array}{lll}
\hat S&=&3\displaystyle\sum_{\mu_1\mu_2}\sum_{\mu'_1\mu'_2}
((\mu'_1\mu'_2\vert (\hat 1_B-\hat T_B)\vert \mu_1\mu_2))
b_{\mu_1}^\dagger b_{\mu_2}^\dagger b_{\mu'_1}b_{\mu'_2}
\\
&&+\displaystyle\sum_{\mu_1\mu_2}\sum_{\mu'_1\mu'_2}
((\mu'_1\mu'_2\vert\hat Z'\vert \mu_1\mu_2))
b_{\mu_1}^\dagger b_{\mu_2}^\dagger b_{\mu'_1}b_{\mu'_2}\text{,}
\end{array}
\end{equation}
Furthermore, if $\hat T_B=\breve 1_B$ holds:
\begin{equation}
\label{eq:sanamd1b}
\begin{array}{lll}
\hat S&=&3\displaystyle\sum_{\mu_1\mu_2}\sum_{\mu'_1\mu'_2}
((\mu'_1\mu'_2\vert (\hat 1_B-\breve 1_B)\vert \mu_1\mu_2))
b_{\mu_1}^\dagger b_{\mu_2}^\dagger b_{\mu'_1}b_{\mu'_2}
\\
&&+\displaystyle\sum_{\mu_1\mu_2}\sum_{\mu'_1\mu'_2}
((\mu'_1\mu'_2\vert\hat Z'\vert \mu_1\mu_2))
b_{\mu_1}^\dagger b_{\mu_2}^\dagger b_{\mu'_1}b_{\mu'_2}\text{,}
\\
&=&3\displaystyle\sum_{\bar t\mu}\sum_{\bar t'\mu'}
((\bar t'\mu'\vert (\hat 1_B-\breve 1_B)\vert \bar t\mu))
b_{\bar t}^\dagger b_{\mu}^\dagger b_{\bar t'}b_{\mu'}
\\
&&+\displaystyle\sum_{\bar t_1\bar t_2}\sum_{\bar t'_1\bar t'_2}
((\bar t'_1\bar t'_2\vert\hat Z'\vert \bar t_1\bar t_2))
b_{\bar t_1}^\dagger b_{\bar t_2}^\dagger b_{\bar t'_1}b_{\bar t'_2}\text{,}
\end{array}
\end{equation}
from which $\hat S\vert N; t)$ can be clearly seen.
From these analyses, it became clear that the incorrect conclusion regarding the Park operator in the conventional Dyson boson expansion method was due to the failure to grasp the conditions satisfied by NAMD, while simultaneously presupposing NAMD.

\section{Discussion}
\label{dis}

The prescription for deriving the Fermion Subspace Mapping (FSSM) from the Whole Fermion Space Mapping (WFSM) has been established by the norm operator method. This implies that constructing a practical boson expansion theory (BET) within the WFSM framework is essentially equivalent to constructing a consistent and practical FSSM. Such a consistent BET has been achieved for the first time by the norm operator method, which overcomes the fundamental flaws of conventional theories.

The ``apparent success" in reproducing large-amplitude collective motions (LACM) using conventional methods---whose logical flaws have now been exposed---does not justify the conclusion that conventional BETs have successfully elucidated the microscopic structure of nuclei. It is imperative to re-examine whether conventional BETs truly reproduced LACM by replacing the conventional formulations with the norm operator method, while maintaining the same Hamiltonian and parameters. The procedure for this verification has already been provided in Section 6, ``Comments on the application to the collective motions of nuclei," of our previous paper \cite{NR23}.

The FSSM mapping via the norm operator method becomes an isomorphic mapping under appropriate conditions. This isomorphism guarantees that the algebraic structure---specifically the commutation relations between phonons adopted as boson excitations, those between adopted and non-adopted phonons, and those between all phonons and scattering operators---is rigorously preserved in the mapped boson operators. Obtaining a correct mapping is difficult if one focuses solely on the reproduction of commutation relations. Indeed, KT-1, which aimed to reproduce commutation relations for all phonon modes, suffered from convergence issues. This led to KT-2, which restricted phonons to collective modes only and was regarded as ``practical'' due to its better convergence. However, as demonstrated in our previous paper \cite{NR23}, NAMD and the small-parameter expansion (where phonons become bosons in the zeroth approximation) are logically incompatible. A primary reason for the flaws in conventional BETs lies in their over-reliance on commutation relations during formulation, which inadvertently provided a ground for accepting NAMD as a "good approximation" without rigorous justification. This misunderstanding was further reinforced by the Dyson BET, which emphasized finite expansions based on NAMD. Such formulations, rooted in a fundamental misunderstanding of NAMD, must be superseded by the norm operator method.

The fact that the FSSM mapping becomes an isomorphic mapping under an appropriate approximation is a crucial key to microscopically elucidating the mechanism by which collective motions in fermion many-body systems emerge as boson excitations. Furthermore, the achievement of the norm operator method---clarifying that the resulting boson expansion becomes substantially a finite expansion when NAMD holds (regardless of whether it is Hermitian or non-Hermitian)---provides a new paradigm for the microscopic foundation of the Interacting Boson Model (IBM) \cite{AI75}, which has been a long-standing challenge.
To date, the Otsuka-Arima-Iachello (OAI) mapping \cite{OAI78} has been considered a promising microscopic foundation for the IBM. However, while the OAI method claims to be a mapping, it does not guarantee an isomorphic mapping.The fact that IBM Hamiltonians consist of a limited number of $s$ and $d$ boson operators suggests that the Hermitian expansion in cases where NAMD holds---a possibility first clarified by the norm operator method---would be one of the promising candidates for the microscopic foundation of the IBM.

The norm operator method gives a robust many-body framework that supports the microscopic theory of large-amplitude collective motions, whose validity must be extensively verified. This method paves the way for further research in nuclear physics to reveal the mechanism by which fermion many-body systems manifest collective motions as boson excitations.

\section{Summary}
\label{sum}

By using the norm operator method, which is a new boson expansion theory (BET), the incorrect claims of conventional BETs regarding the relationship between the whole fermion space mapping (WFSM) and the fermion subspace mapping (FSSM) have been corrected.

Conventional BETs have been formulated by completely ignoring (NAMD: Non-Adopted-Mode-Discretion) the contributions of phonon excitations not adopted as boson excitations in the FSSM, without having the means to confirm the impact of these contributions. Furthermore, the method for obtaining the subspace from the whole space has been limited to simply discarding the boson excitation modes corresponding to the phonon excitation modes that are cut off by NAMD from the physical subspace obtained by the WFSM. It was claimed that the mapping operator of the FSSM could not be obtained from that of the WFSM using this method.
Meanwhile, the following claims were made in the Dyson boson expansion method:
\begin{enumerate}
\item The Park operator, which is used in the WFSM to distinguish whether a boson state vector is a physical state vector, is not effective in the FSSM.
\item This is because the projection operators onto the physical subspace differ between the WFSM and the FSSM, so there is no problem.
\item The phonon and scattering operators obtained by the WFSM, when their boson excitations are limited, match those obtained by the FSSM.
\end{enumerate}

The norm operator method, on the other hand, is formulated without presupposing NAMD.
This method introduces a mapping that, in addition to limiting the types of phonon excitation modes conventionally adopted, also simultaneously limits the number of excitations to obtain a practical boson expansion.
By introducing the norm operator into this mapping operator, it is possible to handle Hermitian and non-Hermitian type mappings comprehensively. Furthermore, by removing the limitations imposed on the phonon excitations, both the WFSM and the conventionally adopted FSSM can be reproduced.
The introduction and utilization of the norm operator allowed us to establish a concrete method for the small parameter expansion (where the phonon becomes a boson in the zeroth approximation), and enabled an accurate understanding of cases where NAMD holds.
Specifically, the following points have been clarified:
\begin{enumerate}
\item The small parameter expansion and NAMD are incompatible.
\item If the small parameter expansion holds, all expansions become infinite expansions regardless of whether they are Hermitian or non-Hermitian types.
\item If NAMD holds, the expansion is substantially a finite expansion.
\end{enumerate}

The analysis using the norm operator method has clarified the following points:
\begin{enumerate}
\item The physical subspace of the WFSM is a complete replica of the entire fermion space. A method to restrict this physical subspace to obtain the FSSM physical space has been established.
\item This method provided a new perspective to the norm operator method: the physical subspace obtained by the FSSM can be derived from the physical subspace of the WFSM by appropriately renormalizing the bosons responsible for the phonon excitations that are discarded by NAMD.
\item When NAMD holds, it was clarified that, contrary to conventional claims, the operator for the FSSM can be derived by limiting the boson excitations of the WFSM mapping operator only to those that remain in NAMD.
\item It was also clarified that when NAMD holds, the Park operator is effective for the FSSM as well, contrary to conventional claims.
\end{enumerate}

We also have discussed how to verify the applicability of the boson expansion theory to large-amplitude collective motions and offered a new perspective on a microscopic foundation of the interacting boson model (IBM).

\let\doi\relax

\end{document}